\documentclass[12pt]{article}
\usepackage{latexsym}
\usepackage{amsfonts}
\usepackage{amsmath}
\begin{document}
\title{ Cartan's equivalence method and null coframes in General Relativity}
\author{Emanuel Gallo\thanks{egallo@fis.uncor.edu} \, Mirta Iriondo\thanks{mirta@fis.uncor.edu} \,
Carlos Kozameh\thanks{kozameh@famaf.unc.edu.ar}\\ \footnotesize{ \textit{FaMAF, Universidad Nacional de
C\'{o}rdoba, 5000 C\'{o}rdoba, Argentina}}}

\date{} \maketitle

\begin{abstract}
Using Cartan's equivalence method for point transformations we
obtain from first principles the conformal geometry associated
with third order ODEs and a special class of PDEs in two
dimensions. We explicitly construct the null tetrads of a family
of Lorentzian metrics, the conformal group in three and four
dimensions and the so called normal metric connection. A special
feature of this connection is that the non vanishing components of
its torsion depend on one relative invariant, the (generalized)
W\"unschmann Invariant. We show that the above mentioned
construction naturally contains the Null Surface Formulation of
General Relativity.

\end{abstract}

\pagenumbering{arabic}
\section{Introduction}

In 1983 a new formulation called Null Surface Formulation of
General Relativity (or NSF for short) presented a radically
different point of view of General Relativity where the emphasis
shifted from a metric tensor on a 4-dim manifold to level surfaces
and PDEs in two dimensions. In NSF, the conformal space time,
i.e., a 4 dimensional manifold equipped with a conformal
structure, arises from solutions of a pair of PDEs in a
2-dimensional space representing the sphere of null
directions~\cite{KN}. A necessary condition for the existence of a
conformal metric is the so called ``metricity condition'', whose
geometric meaning was unknown until recently. The construction of
the 3-dimensional version of NSF, a technically simpler problem
than its higher dimensional analog~\cite{Forni}, proved to be a
key for a deeper understanding of the formalism.  Since this
construction revealed a very strong connection between General
relativity and Cartan's equivalence method, it is very relevant to
briefly review the 3-dimensional model.

Given the following ODE
\begin{equation}
u^{\prime \prime \prime }=F(u,u^{\prime },u^{\prime \prime },s),
\label{third-order1}
\end{equation}
(where the primes means derivatives of the $s$ parameter) it is
easy to see that its solution space is 3-dimensional. Denoting by
$x^a$ the local coordinates in the solution space, the solution to
(\ref{third-order1}) can be written as

\begin{equation}
u=Z(x^a,s), \label{Z}
\end{equation}
Moreover, one can construct the Pfaffian $\theta^i$ associated
with (\ref{third-order1}) as
$$
(\omega^1, \omega^2, \omega^3) \equiv (Z,_a dx^a, Z^{\prime },_a
dx^a,Z^{\prime \prime },_a dx^a)
$$
together with a one-dimensional family of metrics (labelled by
$s$)
$$
g(s) =\theta^1 \theta^3 + \theta^3 \theta^1 - \theta^2\theta^2.
$$
where $\theta^1=\omega^1$, $\theta^2=\omega^2$ and $\theta^3$ is a
linear combination of the $\omega^i$s.

By construction, the level surfaces of (\ref{Z}) are null. Thus,
except for the fact that the induced metric depends on the
parameter $s$, the construction of a space time and its conformal
structure is contained in the solution $Z(x^a,s)$. It can be shown
that when the function $F$ satisfies a differential condition
I[F]=0, the metric $g(s)$ satisfies
$$
\frac{d g(s)}{d s}\propto g(s),
$$
thus giving a unique conformal structure to the space time $x^a$.
The metricity condition I[F]=0 is the 3-dimensional analog of the
much more complicated equation obtained in 4 dimensions.

It was a great surprise to learn that the differential operator $I[F]$ was originally obtained by W\"{u}nchsmann
in 1905, and that the above underlined geometric construction was also previously obtained by E. Cartan in 1942
where he showed that two third order ODE's that are equivalent under a point transformation give rise to two
geometries in the solution space that are isometric~\cite{Cartan}. To obtain these results Cartan introduced a
geometric structure in a 4-dim fiber-space with coordinates $(x^a,s)$ and imposed algebraic and differential
conditions on the so called normal metric connection. Since this approach is completely different from the
standard Equivalence Method, it is useful to show the relationship between Cartan's original approach and the
set of necessary and sufficient conditions that arise when the Equivalence Method is applied to this ODE. It is
also worth mentioning that the normal metric connetion approach was published in a journal of very little impact
and was largely ignored by the mathematical community since it was not further developed for other ODEs or PDEs.
As a result, it is very difficult for the average reader to see why the geometric construction of a so called
normal metric connection is related to a third order ODE modulo point transformations. Only very recently the
full construction of the geometric structure via the Equivalence Method has been completed for third order ODEs
(see \cite{Nur} and section 3) and, as it is done here, the approach can be extended to the much more involved
problem in four dimensions.

Since our main motivation for this work is to understand the four dimensional geometric structure derived from
the same pair of PDEs related to NSF, we apply the Equivalence Method to develop the geometry associated with
these PDEs. The main issues addressed in this work are: the symmetry group associated with the construction, the
explicit construction of the null tetrad, the relative invariants and the associated connection. In particular,
we show that the so called metricity condition is the vanishing of a relative invariant in the Equivalence
Method. Most of these results here presented extend previous results obtained using a different approach
\cite{SCT} providing a straightforward construction with a clear geometrical meaning. It also generalizes recent
work on a torsion free connection with a vanishing generalized W\"{u}nchsmann invariant \cite{GKNP}. This
construction includes a torsion tensor constructed from the relative invariants of the equivalence problem.
However, for historical and pedagogical reasons we briefly review and develop the 3-dimensional before
addressing the technically involved 4-dim version.

This work is divided as follows. In Section 2 we briefly review
Cartan's Equivalence Method applied to ODEs. In Section 3 we apply
this method for 3rd order ODEs that are equivalent under point
transformation constructing the invariants that are needed to
develop the geometric structures. We then relate these results to
the so called Null Surface Formulation of GR in three
dimensions(\cite{Forni,FKNN,SCT}), and show that the metricity
condition is the vanishing of one of the invariants associated to
the equivalence of these third order ODE's. In Section 4 we obtain
the relative invariants from the pair of PDEs that yield
Lorentzian metrics in four dimensions. We then use the invariants
to construct a normal metric connection and show that when the so
called generalized W\"{u}nchsmann invariant vanishes we recover
the standard NSF. Finally, in the Conclusions we summarize our
results and discuss some further work that could be very
interesting to pursue.

\section{Cartan's Equivalence Method}
The so called Cartan's Equivalence Method allows one to find
necessary and sufficient conditions for equivalence
between coframes in n-dimensional manifolds $M$ and $\widetilde{M}$ respectively.(see ~\cite{Olver}).\\
Let $M$ be a manifold of dimension $n$, and $\mathcal{F}(M)$ a
frame bundle over $M$ with structure group $GL(n,\mathbb{R})$. By
a $G$-structure $\mathcal{G}$, we understand a subbundle of
$\mathcal{F}(M)$ over $M$ with structure
group $G\subset GL(n,\mathbb{R})$.\\
Locally we have $\mathcal{G}\simeq M\times G$.\\\\
\textbf{Definition.} \textit{Let $\omega$ and $\widetilde{\omega}$
be coframes in n-dimensional manifolds $M$ and $\widetilde{M}$
respectively. The G-valued equivalence problem for these coframes
is to determine if there exists a local diffeomorphism $\Phi:
M\longrightarrow \widetilde{M}$ and $G$-valued functions
$g:M\longrightarrow G$ and
$\widetilde{g}:\widetilde{M}\longrightarrow G$ such that
$\Phi^*\left [
\widetilde{g}(\widetilde{x})\,\widetilde{\omega}\right
]=g(x)\omega
$.}\\\\
A particular class of equivalence problem comes from n-order
differential equations in $s$ independent and $r$ dependent
variables. Let us concentrate in single $n$-order ODEs. In this
case, $s=r=1$ and the ODE reads
\begin{equation}
u^{(n)}=F\left (s,u,u',...,u^{(n-1)}\right ),\nonumber
\end{equation}
where $s$ is the independent parameter and $u^{(n)}$ denotes the
$n$ derivative of the dependent parameter $u$
with respect to $s$ ($s$ and $u$ take values on sets $\mathbb{X}$ and $\mathbb{U}$ respectively).\\
In this case, $M$ is the $(n-1)$-Jet space
$J^{n-1}\left(\mathbb{X},\mathbb{U}\right )$, with local
coordinates
\\ $\left (s,u,u',...,u^{(n-1)}\right )$, and its associated coframe $\omega$ reads
\begin{eqnarray}
\omega^1&=&du-u'ds,\nonumber\\
\omega^2&=&du'-u''ds,\nonumber\\
&\vdots&\nonumber\\
\omega^{n}&=&du^{(n-1)}-F ds,\nonumber\\
\omega^{(n+1)}&=&ds.\nonumber
\end{eqnarray}
There are three main transformations associated to the equivalence
problem of ODEs:\\\\
$\bullet$ The contact transformations $\Phi:J^1\left
(\mathbb{X},\mathbb{U}\right )\longrightarrow J^1\left
(\mathbb{X},\mathbb{U}\right )$,
$(s,u,u')\rightarrow(\widetilde{s},\widetilde{u},\widetilde{u'})$,
with associated prolongation $p\,^{(n-1)}\Phi$ to $J^{n-1}\left
(\mathbb{X},\mathbb{U}\right )$.\\\\
$\bullet$ The point transformation $\Phi:J^0\left
(\mathbb{X},\mathbb{U}\right )\longrightarrow J^0\left
(\mathbb{X},\mathbb{U}\right )$,
$(s,u)\rightarrow(\widetilde{s},\widetilde{u})$, with associated
prolongation $p\,^{(n-1)}\Phi$ to $J^{n-1}\left
(\mathbb{X},\mathbb{U}\right )$.\\\\
$\bullet$ The fiber preserving transformation which are a subset
of point transformation where the new independent parameter
$\widetilde{s}$ only depends on the old parameter $s$.\\\\
It is easy to show then, that the contact transformation gives
rise to the equivalence problem
$(p\,^{(n-1)}\Phi^*)\widetilde{\theta}=\theta$, with
\begin{equation}
\theta=g\omega=\left(
\begin{array}{cccccc}
  a_1 & 0 & \ldots & \ldots & \ldots & 0 \\
  a_2 & a_3 & 0 & \ldots & \ldots & 0 \\
  a_4 & a_5 & a_6 & 0 & \ldots & 0 \\
  \vdots & \vdots & \vdots & \vdots & \vdots & \vdots \\
   a_{\frac{n^2+n+2}{2}} &  a_{\frac{n^2+n+4}{2}}& 0 & \ldots & \ldots & a_{\frac{n^2+n+6}{2}} \\
\end{array}
\right)\left(
\begin{array}{c}
  \omega^1 \\
  \omega^2 \\
  \vdots \\
  \omega^{n} \\
 \omega^{n+1} \\
\end{array}
\right),
\end{equation}
with all the diagonal components different from zero. In the case
of point transformations, $a_{\frac{n^2+n+4}{2}}=0$, and
in the fiber preserving case we have in addition that $a_{\frac{n^2+n+2}{2}}=0$.\\
Let us compute the $d\theta$ and $d\widetilde{\theta}$.
\begin{eqnarray}
d{\bf \theta}&=& dg\;\wedge{\bf \omega}+g\;d{\bf \omega}\nonumber\\
&=&dg\;g^{-1}\wedge g\;{\bf \omega}+g\;d{\bf \omega}\nonumber\\
&=&\Pi\wedge{\bf \theta}+T_{ij}
\theta^i\wedge\theta^j,\label{dthetag}
\end{eqnarray}
where the coefficients $T_{ij}$ are known as torsion elements, and
$\Pi=dg\,g^{-1}$ is the matrix of Maurer Cartan forms $\pi^A$,
which can be written as $\Pi^i_k=C^i_{kA}\pi^A$, with $C^i_{kA}$
constant coefficients. In this notation small case indices run
from $1$ to $n$ while capital indices run through the group
dimension. Using index notation we can write (\ref{dthetag}) as
\begin{eqnarray}
d{ \theta^i}&=&C^i_{kA}\pi^A\wedge{\theta^k}+T^i_{jk}
\theta^j\wedge\theta^k,\label{dtheta}
\end{eqnarray}
 In a similar way, we have
\begin{eqnarray}
d{\widetilde{\theta}^i}&=&C^i_{kA}\widetilde{\pi}^A\wedge{
\widetilde{\theta}^k}+\widetilde{T}^i_{jk}
\widetilde{\theta}^j\wedge\widetilde{\theta}^k.\label{dthetabbar}
\end{eqnarray}
Note that we write $C^i_{kA}$ instead of $\widetilde{C}^i_{kA}$
since the group parameters enter in identical
way in $g$ and $\widetilde{g}$.\\

The idea in the Equivalence Method is to obtain as many group
coefficients $g$ ($\widetilde{g}$) functionally dependent on the
associated Pfaffian system $\omega$ ($\widetilde{\omega}$) to
obtain what is called a rigid co-frame where the solution to the
equivalence problem is straightforward~\cite{Olver}.

 The first step in the solution of the equivalence problem is to note that if $\widetilde{\theta}=\theta$,
 then $d\widetilde{\theta}=d\theta$,  where we have omitted the pullbacks for simplicity. It then follows that
\begin{equation}
\left [C^i_{kA}(\pi^A-\widetilde{\pi}^A)+\left
(T^i_{jk}-\widetilde{T}^i_{jk}\right )\theta^j\right
]\wedge\theta^k=0.\label{lemma}
\end{equation}
Furthermore, due to a Cartan lemma, there exist functions
$f^i_{kj}=f^i_{jk}$ such that
\begin{equation}
\left [C^i_{kA}(\pi^A-\widetilde{\pi}^A)+\left
(T^i_{jk}-\widetilde{T}^i_{jk}\right )\theta^j\right
]=f^i_{kj}\theta^j.
\end{equation}
The above equation implies that there exist functions
$\lambda^A_k$ such that
\begin{equation}
\widetilde{\pi}^A=\pi^A+\lambda^A_k\theta^k\label{pitrans},
\end{equation}
and
\begin{equation}
\widetilde{T}^i_{jk}=T^i_{jk}+C^i_{kA}\lambda^A_j-C^i_{jA}\lambda^A_k\label{tortrans}.
\end{equation}
Eqs. (\ref{pitrans}) and (\ref{tortrans}) can be used to eliminate
the maximal possible number of torsion components
$\widetilde{T}^i_{jk}$. This technique, one of the fundamentals in
Cartan's equivalence method, is called absorbtion of the torsion.
Let's suppose that no more absorbtion is possible. Then
 (\ref{dtheta}) reads
\begin{equation}
d\theta^i=C^i_{kA}\pi^A\wedge\theta^k+U^i_{jk}
\theta^j\wedge\theta^k,
\end{equation}
where the ${\pi^A}$ are the new redefined Maurer-Cartan forms
modulo $\theta^i$, and $U^i_{jk}=U^i_{jk}(x,g)$, are the non
absorbable components of the torsion known as essential torsion.
This name is justified if we note that after absorbtion, they are
linearly independent from the Maurer-Cartan forms, and thus,
\begin{equation}
(p\,^{n-1}\Phi^*)\widetilde{U}^i_{jk}(\widetilde{x},\widetilde{g})=U^i_{jk}(x,g)\label{invariants}.
\end{equation}
Then the essential torsion components are invariants of the
problem for any choice of the group parameters.

The second technique in Cartan's algorithm is known as
normalization. If possible, we can normalize each invariant to a
constant (generally zero or one) and use this procedure to
eliminate one of the group parameters. We then reduce the
structure group by the number of ``eliminated'' parameters and
obtain a new \textit{normalized} coframe. This procedure is called
a loop in Cartan's method. If the invariants (\ref{invariants}) do
not explicitly depend on the group parameters, they are true
invariants of the problem, and they provide necessary conditions
for the equivalence problem of two given coframes. We can continue
applying the absorbtion and normalization techniques to the
normalized coframe until we reach one of two possible scenarios.
In the first case, we can successfully normalize all group
parameters, and the problem is reduced to a $\{e\}$-structure (i.e
a structure where $G$ consist in the identity). We have obtained
an invariant coframe, and it can be shown (see \cite{Olver}) that
from the true invariants and its derivatives, we can construct a
maximal set of functionally independent invariants which form a
set of
necessary and sufficient conditions for the equivalence problem.\\
In the second case, after a finite number of loops in the Cartan's
method, we end up with a system where some group parameters are
not determined, although no more normalization is possible. In
this case, we must use the so called Cartan test of
involution~\cite{Olver}. This test decides if the equivalence
problem in question has an underlying infinite symmetry group (we
say then that the system is in involution), or if the problem has
a finite symmetry group. In the last case, we can apply the method
known as prolongation.

Let´s assume that we are able to determine all new $\pi^A$ modulo
some $\lambda^A_j$, i.e., the $\widetilde{\pi^A}$ can be written
as
\begin{equation} \widetilde{\pi}^A=\pi^A+\lambda^A_{Dj}\theta^j
+\lambda^A_{Fj}\theta^j,
\end{equation}
where $\lambda^A_{Dj}$ are determined by the absorbtion procedure
and the $\lambda^A_{Fj}$ (free
functions) are not determined by the procedure.\\
 It can then be shown that to solve the original problem, namely to find the symmetry group together with
 the maximal set of invariants associated with the equivalence problem, it is equivalent to solve a problem
 where,

 \begin{enumerate}
\item the free parameters of $G$ become coordinates of an enlarged
base space  $M^{(1)}=M\times G$, and \item the free functions
$\lambda^A_{Fj}$ become parameters of an enlarged group $G^{(1)}$.
\end{enumerate}

We first extend the original coframe by including the new forms
\begin{equation}
\kappa^A=\pi^A+\lambda^A_{Dj}\theta^j,
\end{equation}
On the base space $M^{(1)}$ we consider the system
${\theta^i,\kappa^A}$, with the structure group given by
\begin{equation}
G^{(1)}=\left(%
\begin{array}{cc}
  \mathbb{I} & 0 \\
  \lambda^A_{Fj} & \mathbb{I} \\
\end{array}%
\right),
\end{equation}
where $\mathbb{I}$ is the identity on the respective space.\\
If we study this prolonged problem we can, in principle, normalize
some or all free functions, (maybe applying the prolongation more
than once), and in this way we can find all the necessary and
sufficient invariants of the equivalence problem. For more details
and applications of this method we refer to the Olver
book~\cite{Olver}.

\section{The third order ODE}
In this section we apply Cartan's equivalence method to find the
class of equivalent third order ODEs under point transformations.
This is a non trivial problem and the set of necessary and
sufficient conditions for such a class have been obtained recently
by P. Nurowski~\cite{Nur_nuevo}. (A similar question for fibre
preserving or contact transformations is also available in the
literature(\cite{Nur,Neut,Ch,Grebot}). The results given in this
section have been derived independently and in a more restricted
context to show the explicit steps that
will later generalize to 4 dimensions.\\
 We will say that the equation
\begin{equation}
u^{\prime \prime \prime }=F(u,u^{\prime },u^{\prime \prime
},s),\label{third}
\end{equation}
is equivalent to
\begin{equation}
\widetilde{u}^{\prime \prime \prime
}=\widetilde{F}(\widetilde{u},\widetilde{u}^{\prime
},\widetilde{u}^{\prime \prime },\widetilde{s}),
\end{equation}
if there exist a point transformation
%$\Phi:J^0(\mathbb{R},\mathbb{R})\longrightarrow
%J^0(\mathbb{R},\mathbb{R}),$
\begin{eqnarray}
%\widetilde{s} \circ \Phi=\xi(s,u),\\
%\widetilde{u} \circ \Phi=\psi(s,u),
\widetilde{s}=\xi(s,u),\\
\widetilde{u}=\psi(s,u),
\end{eqnarray}
with prolongation
%$p\,^2\Phi: J^2(\mathbb{R},\mathbb{R})\longrightarrow
%J^2(\mathbb{R},\mathbb{R})$,
\begin{equation}
(s,u,w,r)\longrightarrow
(\widetilde{s},\widetilde{u},\widetilde{w},\widetilde{r})=\left
(\xi(s,u),\psi(s,u),\frac{\psi_s+w\psi_u}{\xi_s+w\xi_u},\frac{w\widetilde{w}_u+\widetilde{w}_s+r\widetilde{w}_w}{w\xi_u+\xi_s}
\right )\nonumber
\end{equation}
which transforms one equation into the other. Note that we have
made use of the following notation
\begin{equation}
(s,u,u',u'')= (s,u,w,r),\qquad \mbox{and}\qquad
(\widetilde{s},\widetilde{u},\widetilde{u}',\widetilde{u}'')=
(\widetilde{s},\widetilde{u},\widetilde{w},\widetilde{r}),
\end{equation}
to label the coordinates of $J^2(\mathbb{R},\mathbb{R})$. When
appropriate,  we will also define
\begin{eqnarray}
{\bf x}&=&(s,u,w,r),\nonumber\\
{\bf \widetilde{x}}&=&(\widetilde{s},\widetilde{u},\widetilde{w},\widetilde{r}),\nonumber\\
{\bf\omega}&=&(\omega^1,\omega^2,\omega^3,\omega^4),\nonumber\\
{\bf\theta}&=&(\theta^1,\theta^2,\theta^3,\theta^4).\nonumber
\end{eqnarray}
The Pfaffian system $\mathcal{P}$ associated to the equation
(\ref{third}) is
\begin{eqnarray}
\omega^1&=&du-w\; ds,\\
\omega^2&=&dw-r\; ds,\label{omega}\\
\omega^3&=&dr- F\; ds,\\
\omega^4&=&ds,
\end{eqnarray}
and local solutions of (\ref{third}) are in correspondence one to
one with integral curves $\gamma: \mathbb{R}\longrightarrow
J^2(\mathbb{R},\mathbb{R})$ of $\mathcal{P}$ satisfying
$\gamma^*\,ds\neq0$. These curves are generated by the vector
field on $J^2(\mathbb{R},\mathbb{R})$ given by
\begin{equation}
e_s=D=\frac{\partial}{\partial s}+w\frac{\partial}{\partial
u}+r\frac{\partial}{\partial w}+F\frac{\partial}{\partial r}.
\end{equation}
We will restrict the domain of definition of $F$ to a open
neighborhood $U$ of $J^2(\mathbb{R},\mathbb{R})$ where $F$ is
$C^\infty$ and the Cauchy problem is well posed.
 Then, it follows from Frobenius's theorem that the solution space $M$ is a three dimensional $C^\infty$ manifold,
 parametrized by the integration constants $x^a=(x^1,x^2,x^3)$.\\
The solution of eq. (\ref{third}), $u=Z(s,x^a)$, induces a
diffeomorphism
 $\zeta: M\times\mathbb{R} \longrightarrow
J^2(\mathbb{R},\mathbb{R})$ given by $(s,x^a)\longrightarrow
(s,Z,Z',Z'')$. On $M\times\mathbb{R}$ the pullback of the Pfaffian
forms $w^i$ is given by
\begin{eqnarray}
\beta^1&=&Z_a\,dx^a,\nonumber\\
\beta^2&=&Z'_a\,dx^a,\nonumber\\
\beta^3&=&Z''_a\,dx^a,\nonumber\\
\beta^4&=&ds.\nonumber
\end{eqnarray}

We now study the equivalence problem of (\ref{third}) under point
transformations $\Phi:J^0\rightarrow J^0$. This problem gives the
following equivalent problem of $G$-structures,
($p\,^2\Phi^*)\,\bf {\widetilde{\theta}}=\bf{\theta}$, with
\begin{equation}
\left (\begin{array}{c}
\theta^1\\
\theta^2\\
\theta^3\\
\theta^4
\end{array}
\right )=\left (\begin{array}{cccc}
a_1&0&0&0\\
a_2&a_3&0&0\\
a_4&a_5&a_6&0\\
a_7&0&0&a_8
\end{array}
\right )\left (\begin{array}{c}
\omega^1\\
\omega^2\\
\omega^3\\
\omega^4
\end{array}
\right ), \label{g-estructura}
\end{equation}
and a similar expression for $\widetilde{\theta}$. In compact
notation we write (\ref{g-estructura}) as $\theta=g\omega$.\\
Differentiating ${\bf \theta}$, we obtain
\begin{eqnarray}
d{\bf \theta}&=& dg\;\wedge{\bf \omega}+g\;d{\bf \omega}\\
&=&dg\;g^{-1}\wedge g\;{\bf \omega}+g\;d{\bf \omega}\\
&=&\Pi\wedge{\bf \theta}+T_{ij} \theta^i\wedge\theta^j,
\end{eqnarray}
where $T_{ij} \theta^i\wedge\theta^j=g\;d{\bf \omega}$ and
$$
\Pi=dg\;g^{-1}=\left (\begin{array}{cccc}
\pi^1&0&0&0\\
\pi^2&\pi^3&0&0\\
\pi^4&\pi^5&\pi^6&0\\
\pi^7&0&0&\pi^8
\end{array}
\right ),
$$
with
$$
\pi^1=\frac{da_1}{a_1},\; \pi^2=\frac{da_2}{a_1}-\frac{da_3
a_2}{a_1 a_3}, \;\pi^3=\frac{da_3}{a_3},
$$
$$\pi^4=\frac{da_4}{a_1}-\frac{da_5 a_2}{a_1 a_3}-\frac{da_6 (-a_2 a_5+a_4 a_3)}{a_1 a_3 a_6},
$$
$$
\pi^5=\frac{da_5}{a_3}-\frac{da_6 a_5}{a_3 a_6}, \; \pi^6=\frac{da_6}{a_6}, \; \pi^7=\frac{da_7}{a_1}-\frac{da_8
a_7}{a_1 a_8},\;  \pi^8=\frac{da_8}{a_8}.$$

{\bf First Loop}: From ${\bf \omega}=g^{-1} {\bf \theta}$, we obtain the following structure equations
\begin{eqnarray}
d\theta^1&=&\pi^1\wedge\theta^1+T^1_{24}\theta^2\wedge\theta^4+T^1_{21}\theta^2\wedge\theta^1+T^1_{14}\theta^1\wedge\theta^4,\label{theta1}\\
d\theta^2&=&\pi^2\wedge\theta^1+\pi^3\wedge\theta^2+T^2_{24}\theta^2\wedge\theta^4+T^2_{21}\theta^2\wedge\theta^1\nonumber\\
&&+T^2_{14}\theta^1\wedge\theta^4
+T^2_{34}\theta^3\wedge\theta^4+T^2_{31}\theta^3\wedge\theta^1,\label{theta2}\\
d\theta^3&=&\pi^4\wedge\theta^1+\pi^5\wedge\theta^2+\pi^6\wedge\theta^3+T^3_{34}\theta^3\wedge\theta^4\nonumber\\
&&+T^3_{21}\theta^2\wedge\theta^1+T^3_{14}\theta^1\wedge\theta^4
+T^3_{24}\theta^2\wedge\theta^4+T^3_{31}\theta^3\wedge\theta^1,\label{theta3}\\
d\theta^4&=&\pi^7\wedge\theta^1+\pi^8\wedge\theta^4+T^4_{24}\theta^2\wedge\theta^4+T^4_{21}
\theta^2\wedge\theta^1+T^4_{14}\theta^1\wedge\theta^4.\label{theta4}
\end{eqnarray}
Using the freedom $ \pi^A\to\pi^A+\lambda^A_j \theta^j,$ many
torsion coefficients can be absorbed. For example, choosing
$\lambda^1_2 = - T^1_{21}$ and  $\lambda^1_4 = T^1_{14}$ it is
easy to see that $\widetilde{T}^1_{21}=\widetilde{T}^1_{14}=0$.
Omitting the $\widetilde{}\;$ for simplicity, the new equations
read
\begin{eqnarray}
d\theta^1&=&\pi^1\wedge\theta^1+T^1_{24}\theta^2\wedge\theta^4,\label{theta1a}\\
d\theta^2&=&\pi^2\wedge\theta^1+\pi^3\wedge\theta^2
+T^2_{34}\theta^3\wedge\theta^4,\label{theta2a}\\
d\theta^3&=&\pi^4\wedge\theta^1+\pi^5\wedge\theta^2+\pi^6\wedge\theta^3,\label{theta3a}\\
d\theta^4&=&\pi^7\wedge\theta^1+\pi^8\wedge\theta^4.\label{theta4a}
\end{eqnarray}
with $\displaystyle{T^1_{24}=-\frac{a_1}{a_3 a_8}}$ and $\displaystyle{T^2_{34}=-\frac{a_3}{a_8 a_6}}$.
Normalizing $\displaystyle{T^1_{24}=-1}$ and $\displaystyle{T^2_{34}=-1}$, we determine $a_6$ and $a_8$. The
matrices $g$ and $\Pi$ become

\begin{equation}
g=\left (\begin{array}{cccc}
a_1&0&0&0\\
a_2&a_3&0&0\\
a_4&a_5&\displaystyle{\frac{a_3^2}{a_1}}&0\\
a_7&0&0&\displaystyle{\frac{a_1}{a_3}}
\end{array}
\right ),\label{g}
\end{equation}

$$
\Pi=\left (\begin{array}{cccc}
\pi^1&0&0&0\\
\pi^2&\pi^3&0&0\\
\pi^4&\pi^5&2\pi^3-\pi^1&0\\
\pi^7&0&0&\pi^1-\pi^3
\end{array}
\right ).
$$
{\bf Second Loop:} With the new matrix $g$ we compute again the structure equations. Using the freedom in
$\pi^i$ to eliminate many torsion coefficients we obtain,

\begin{eqnarray}
d\theta^1&=&\pi^1\wedge\theta^1-\theta^2\wedge\theta^4,\label{2theta1a}\\
d\theta^2&=&\pi^2\wedge\theta^1+\pi^3\wedge\theta^2
-\theta^3\wedge\theta^4,\label{2theta2a}\\
d\theta^3&=&\pi^4\wedge\theta^1+\pi^5\wedge\theta^2+(2\pi^3-\pi^1)\wedge\theta^3+T^3_{34} \theta^3\wedge\theta^4,\label{2theta3a}\\
d\theta^4&=&\pi^7\wedge\theta^1+(\pi^1-\pi^3)\wedge\theta^4,\label{2theta4a}
\end{eqnarray}
with $\displaystyle{T^3_{34}=-\frac{3 a_5 a_1+a_3^2
F_r-3a_2a_3}{a_3 a_1}}$. Normalizing the invariant
$\displaystyle{T^3_{34}=0}$ we determine $a_5$. The matrices $g$
and $\Pi$ become \begin{equation} g=\left (\begin{array}{cccc}
a_1&0&0&0\\
a_2&a_3&0&0\\
a_4&\displaystyle{\frac{a_3 a_2}{a_1}-\frac{a_3^2 F_r}{3a_1}}&\displaystyle{\frac{a_3^2}{a_1}}&0\\
a_7&0&0&\displaystyle{\frac{a_1}{a_3}}
\end{array}
\right ),\label{S}
\end{equation}
and
\begin{equation}
\Pi=\left (\begin{array}{cccc}
\pi^1&0&0&0\\
\pi^2&\pi^3&0&0\\
\pi^4&\displaystyle{\pi^2-\frac{a_3d(F_r)}{a_1}}&2\pi^3-\pi^1&0\\
\pi^7&0&0&\pi^1-\pi^3
\end{array}
\right ).\label{pi} \end{equation}

{\bf Third Loop:} Applying Cartan's method one more time and after a new absorption of the Torsion we get

\begin{eqnarray}
d\theta^1&=&\pi^1\wedge\theta^1-\theta^2\wedge\theta^4,\label{3theta1a}\\
d\theta^2&=&\pi^2\wedge\theta^1+\pi^3\wedge\theta^2
-\theta^3\wedge\theta^4,\label{3theta2a}\\
d\theta^3&=&\pi^4\wedge\theta^1+\pi^2\wedge\theta^2+(2\pi^3-\pi^1)\wedge\theta^3+T^3_{24} \theta^2\wedge\theta^4,\label{3theta3a}\\
d\theta^4&=&\pi^7\wedge\theta^1+(\pi^1-\pi^3)\wedge\theta^4+T^4_{24}\theta^2\wedge\theta^4,\label{3theta4a}
\end{eqnarray}
with $\displaystyle{T^3_{24}=-\frac{(2a_3^2F_r^2-9a_2^2+18a_4a_1-3a_3^2DF_r+9F_wa_3^2)}{9a_1^2}}$,
$\displaystyle{T^4_{24}=-\frac{(6a_7a_3-F_{rr}a_1)}{6a_3a_1}}$.\\

Normalizing $T^3_{24}=T^4_{24}=0$ the elements $a_4$ and $a_7$ in
matrices $g$ and $\Pi$ in equations (\ref{S}) and $\pi^4$ and
$\pi^7$ in (\ref{pi}) become
\begin{eqnarray*}
a_4&=&\frac{a_3^2 DF_r}{6 a_1}-\frac{F_w a_3^2}{2 a_1}-\frac{F_r^2 a_3^2}{9 a_1}+\frac{a_2^2}{2 a_1},\\
a_7 &=& \frac{a_1 F_{rr}}{a_3},\\
\pi^4&=&-\frac{(2 a_3 F_r-3 a_2) a_3 d(F_r)}{ 9 a_1^2}-\frac{a_3^2 d(F_w)}{2 a_1^2}+\frac{a_3^2 d(DF_r)}{6 a_1^2},\\
\pi^7&=&\frac{d(F_{rr})}{6 a_3}.
\end{eqnarray*}
{\bf Fourth Loop:} After absorbing the non essential components of the torsion the $d\theta^i$ read

\begin{eqnarray}
d\theta^1&=&\pi^1\wedge\theta^1-\theta^2\wedge\theta^4,\label{4theta1a}\\
d\theta^2&=&\pi^2\wedge\theta^1+\pi^3\wedge\theta^2
-\theta^3\wedge\theta^4,\label{4theta2a}\\
d\theta^3&=&\pi^2\wedge\theta^2+(2\pi^3-\pi^1)\wedge\theta^3+I_1\theta^1\wedge\theta^4,\label{4theta3a}\\
d\theta^4&=&(\pi^1-\pi^3)\wedge\theta^4+I_2\theta^2\wedge\theta^1+I_3\theta^3\wedge\theta^1,\label{4theta4a}
\end{eqnarray}
where
\begin{eqnarray}
I_1&=&\frac{a_3^3}{a_1^3}\left (F_u-\frac{F_r DF_r}{3}+\frac{F_r
F_w}{3}+\frac{2 F_r^3}{27}-\frac{DF_w}{2}+\frac{D^2 F_r}{6}\right
),\\
 I_2&=&\frac{1}{a_3^2} \left (F_{rrw}+\frac{F_{rrr}F_r}{3}+\frac{F_{rr}^2}{6}\right )- \frac{a_2}{
 a_1}I_3,\\
 I_3&=&\frac{a_1}{6a_3^3}F_{rrr}.
\end{eqnarray}
So far, we have three invariants whose vanishing do not depend on
the group parameters. To solve the equivalence problem one must
study different branches of the problem, i.e. different possible
values of the invariants. One then follows a procedure called
prolongation to find a maximal set of invariants which univocally
characterize the equivalence problem. This problem was solved
in~\cite{Nur_nuevo}, obtaining

\begin{eqnarray}
d\theta^1&=&\pi^1\wedge\theta^1-\theta^2\wedge\theta^4,\nonumber\\
d\theta^2&=&\pi^2\wedge\theta^1+\pi^3\wedge\theta^2
-\theta^3\wedge\theta^4,\nonumber\\
d\theta^3&=&\pi^2\wedge\theta^2+(2\pi^3-\pi^1)\wedge\theta^3+I_1\theta^1\wedge\theta^4,\nonumber\\
d\theta^4&=&(\pi^1-\pi^3)\wedge\theta^4+I_2\theta^2\wedge\theta^1+I_3\theta^3\wedge\theta^1,\nonumber\\
d\pi^1&=&-\pi^2\wedge\theta^4+I_4 \,\theta^1\wedge\theta^2+I_5  \,\theta^1\wedge\theta^3+I_6 \, \theta^1\wedge\theta^4-I_3  \,\theta^2\wedge\theta^3,\label{Nurowski}\\
d\pi^2&=&(\pi^3-\pi^1)\wedge\pi^2+I_7 \, \theta^1\wedge\theta^2+I_8 \,\theta^1\wedge\theta^3+I_9 \,\theta^1\wedge\theta^4\nonumber\\
&&+I_{10} \,\theta^2\wedge\theta^3+I_{11} \,\theta^2\wedge\theta^4,\nonumber\\
d\pi^3&=&\frac{I_8+I_4}{2} \,\theta^1\wedge\theta^2+2(I_5-I_{10}) \,\theta^1\wedge\theta^3+I_{11} \,\theta^1\wedge\theta^4\nonumber\\
&&-2I_3 \,\theta^2\wedge\theta^3.\nonumber
\end{eqnarray}
with $I_1$ - $I_{11}$ explicit functionals of $F$.
\subsection{The normal metric connection}
Following Cartan~\cite{Cartan}, in this section we introduce geometrical structures that are naturally induced
by the Pfaffian system associated with the 3rd. order ODE. These new structures give us an alternative method to
show equivalence between ODEs that are related by a point transformation. We emphasize again that this method is
apparently very different from the so called Equivalence Method. Only at the end of the section we show that
both methods are equivalent.

 We first introduce a class of metrics on the solution space and then a generalized  connection with
certain conditions imposed on its torsion and curvature

We recall that after completion of the four loops we have the following pfaffian forms equivalent to the
original:
\begin{eqnarray}
\theta^1&=&a_1\omega^1,\label{pfaff1}\\
\theta^2&=&a_2\omega^1+a_3\omega^2,\\
\theta^3&=&\left (\frac{a_2^2}{2a_1}+\frac{a_3^2}{a_1}a\right
)\omega^1+
\left (\frac{a_3a_2}{a_1}+\frac{a_3^2}{a_1}b\right )\omega^2+\frac{a_3^2}{a_1}\omega^3,\\
\theta^4&=&\frac{a_1}{a_3}\left (c\omega^1+\omega^4\right
),\label{pfaff4}
\end{eqnarray}
with
$$
a=-\frac{1}{2}F_{w}-\frac{1}{9}F^2_{r}+\frac{1}{6}D F_{r},\; \; \; \; b=-\frac{1}{3}F_{r}, \; \; \; \;
c=\frac{1}{6}F_{rr},
$$
and where $a_1,a_2,a_3$ are arbitrary functions on $J^2(\mathbb{R},\mathbb{R})$.\\
Following Cartan, we define the following basis
\begin{eqnarray}
\theta^1_c&=&\omega^1,\\
\theta^2_c&=&\omega^2,\\
\theta^3_c&=&a\,\omega^1+b\,\omega^2+\omega^3,\\
\theta^4_c&=&c\,\omega^1+\omega^4.
\end{eqnarray}
Note that this basis only depends on the associated third order
ODE. In this sense it is an invariant basis with respect to the
subgroup of G with parameters $a_1$, $a_2$, and $a_3$.

The forms (\ref{pfaff1}-\ref{pfaff4}) can be written as
\begin{eqnarray}
\theta^1&=&a_1\,\theta^1_c,\\
\theta^2&=&a_3\left (\theta^2_c+\frac{a_2}{a_3}\,\theta^1_c\right),\\
\theta^3&=&\frac{a_3^2}{a_1}\left [\frac{1}{2}{\left (\frac{a_2}{a_3}\right )}^2\,\theta^1_c+\frac{a_2}{a_3}\,\theta^2_c+\theta^3_c \right ],\\
\theta^4&=&\frac{a_1}{a_3}\,\theta^4_c.
\end{eqnarray}
Using the three 1-forms $\theta^1,\theta^2,\theta^3$ we can
construct a quadratic differential form on
$J^2(\mathbb{R},\mathbb{R})$,
\begin{equation}
h(\textbf{x})=2\theta^{(1}\otimes\theta^{3)}-\theta^2\otimes\theta^2=\eta_{ij}\theta^i\otimes\theta^j,
\end{equation}
where
$$
\eta_{ij}=\left (\begin{array}{ccc}
0&0&1\\
0&-1&0\\
1&0&0\\
\end{array}
\right ).$$

The map $\zeta: M\times\mathbb{R} \longrightarrow
J^2(\mathbb{R},\mathbb{R})$ discussed in section 3 introduces a
quadratic form on $M\times\mathbb{R}$, namely
\begin{equation}
h(x^a,s)=\zeta^*\,h. \end{equation}

This form can be interpreted as a 1-parameter ($s$ being the
parameter) family of lorentzian conformal metrics on $M$.
Furthermore, defining
\begin{equation}
h_c(x^a,s)=\zeta^* \left ( \eta_{ij}\theta^i_c\otimes\theta^j_c
\right ),
\end{equation}
we have
\begin{equation}
h(x^a,s)=\zeta^*\,[a_3^2\,h_c({\bf x})]=\Omega^2\,h_c(x^a,s).
\end{equation}
Hence, we can interpret the $\zeta^*\,\theta^i$ as a family of
null triads which live in the solution space $M$ associated to the
3rd order ODE's. Moreover, we can see that $a_1,a_2,a_3$ are
parameters in a group $G$ which plays
a similar role as in the conformal Lorentz group $CO(2,1)$:\\\\
$\bullet$\; $a_1$  play the role of a  boost $\lambda$ in the
direction of the null vector $e_1$ dual to
$\theta^1$.\\
$\bullet$\; $\frac{a_2}{a_3}$ is a null rotation $\gamma$
around $e_1$.\\
$\bullet$\; $a_3$ is a conformal factor $\Omega$ applied to the triad.\\\\
Later we will see that if $I_1=0$, then we will have that
$G=CO(2,1)$, with the parameter $s$ as a space-like rotation
applied to the triads which make all metrics in the family
conformal to each
other.\\\\
\textbf{Remark 1}  \textit{From now on, to simplify the
expressions, we will not write the pull-back $\zeta^*$.
For example we will write $\theta^i$ instead of $\zeta^*\,\theta^i$.}\\

So far we have constructed a 1-parameter family of conformal
metrics on the solution space of a third order ODE. Let us go
beyond, and add more geometrical structures compatible with this
conformal family of metrics. In particular, we will define in a
univocal way a generalized connection on
$J^2(\mathbb{R},\mathbb{R})$ associated with the null forms
$\theta^i$ with $i=1,2,3$ which characterizes the equivalence of
third order ODEs under point transformations.
This generalized connection satisfies three conditions:\\\\
I) It is a Weyl connection, i.e:
\begin{equation}
\omega_{ij}=\eta_{ik}\omega^k_j=\omega_{[ij]}+\eta_{ij}A,
\end{equation}
with $A$ a 1-form
\begin{equation}
A=A_i\,\theta^i+A_4\,\theta^4.
\end{equation}\\
II) Its associated torsion has a vanishing projection on the base space (with coordinates $x^a$).\\\\
III) Its fiber part (coordinatized by the parameter $s$) only
depends on the nontrivial invariants of the equivalence
problem, i.e.,\\
\begin{eqnarray}
T^1&=&d\theta^1+\omega^1_j\wedge\theta^j=0,\\
T^2&=&d\theta^2+\omega^2_j\wedge\theta^j=0,\\
T^3&=&d\theta^3+\omega^3_j\wedge\theta^j=I_1\,\theta^1\wedge\theta^4
\end{eqnarray}

The reader should distinguish between the torsion of the
connection introduced in the above equations and the torsion
coefficients defined in the previous sections. They have no
relationship at all but are given identical
names in the mathematical literature.\\\\
\noindent\textbf{Remark 2} \textit{Note that the $\omega^i_j$ have components in all $\theta^A$, i.e.:}\\
\begin{equation}
\omega^i_j=\omega^i_{jh}\theta^h+\omega^i_{j4}\theta^4.
\end{equation}
\textit{Thus, it is not a standard connection defined on $M$.}\\\\
\noindent\textbf{Remark 3} \textit{The invariants can be written
in terms of $a,b,c$ as follow:}
\begin{eqnarray}
I_1&=&-\frac{a_3^3}{a_1^3} \left (F_u+2ab+ D a\right ),\\
I_2&=&\frac{1}{a_3^2}\left (c_{w}-c_{r}b+c^2\right )-\frac{a_2}{a_1}I_3,\\
I_3&=&\frac{a_1}{a_3^3}\,c_{r}.
\end{eqnarray}
\textit{In fact they can be written only in terms of $a$ and $b$,
but using $a,b$ and $c$, the expressions
look more compact.}\\\\
Since $a_1,a_2,a_3$ parameterize the group $G$, the connection can
be written modulo a gauge induced by these group. If we take
$a_2=0$ and $a_1=a_3=1$, then the correspondent connection one
form will be denoted by $\widetilde{\omega}$. If we want to write
the connection in another gauge, we use
\begin{equation}
\omega=g\,\widetilde{\omega}\,g^{-1}+dg\,g^{-1}, \end{equation}
where $g$ is an
element of $G$.\\
In this gauge, the invariants reads:
\begin{eqnarray}
I_1&=&-\left (F_u+2ab+D a\right ),\\
I_2&=&c_{w}-c_{r}b+c^2,\\
I_3&=&c_{r}.
\end{eqnarray}
The connection that satisfies  the three conditions is given by,
\begin{eqnarray}
\widetilde{\omega}_{[12]}&=&(-b_u-3ca+a_{w}-a_{r}b)\,\theta_c^1+(cb+A_1)\,\theta_c^2+(c+A_2)\,\theta_c^3+a\,\theta_c^4,\nonumber\\
\widetilde{\omega}_{[13]}&=&(a_{r}-2cb-A_1)\,\theta_c^1-c\,\theta_c^2+A_3\,\theta_c^3+b\,\theta_c^4\nonumber,\\
\widetilde{\omega}_{[23]}&=&(-2c-A_2)\,\theta_c^1-A_3\,\theta_c^2+\theta_c^4\nonumber,\\
\label{omega23}
A&=&A_1\,\theta_c^1+A_2\,\theta_c^2+A_3\,\theta_c^3+b\,\theta_c^4.
\end{eqnarray}
Note that the space components of $A$ are still undetermined and
that the remaining invariants $I_2,I_3$ (which appear in
$d\theta^4$) still need to be included in the geometry. From the
above expression for $\widetilde{\omega}_{[ij]}$ it is easy to see
that the corresponding curvature 2-form
$$\widetilde{\Omega}_{ij}=d\widetilde{\omega}_{ij}+\widetilde{\omega}_{ik}\wedge\widetilde{\omega}^k_{j},$$
will include $I_2,I_3$. Thus, if we impose
\begin{eqnarray}
\widetilde{\Omega}_{23}&=&d\widetilde{\omega}_{23}+\eta^{3i}\widetilde{\omega}_{ih}\wedge\widetilde{\omega}_{h2}\\
&=&I_2\,\theta_c^2\wedge\theta_c^1+I_3\,\theta_c^3\wedge\theta_c^1,\label{curvatura}
\end{eqnarray}
we immediately obtain $ A_1= Dc-2cb+a_r, A_2=-2c, A_3=0.$

In summary, we have constructed in a univocal way a connection
$\omega_{ij}$ such that
\begin{eqnarray}
T^1&=&0,\\
T^2&=&0,\\
T_3&=&I_1\,\theta_c^1\wedge\theta_c^4,\\
\widetilde{\Omega}_{23}&=&I_2\,\theta_c^2\wedge\theta_c^1+I_3\,\theta_c^3\wedge\theta_c^1.
\end{eqnarray}
The skew part of this connection reads,
\begin{eqnarray}
\widetilde{\omega}_{[12]}&=&\left
(-3ca+a_w-a_rb-b_u\right)\theta_c^1+\left (-cb+Dc+a_r\right
)\theta_c^2-c\,\theta_c^3+a\,\theta_c^4,\nonumber\\
\widetilde{\omega}_{[13]}&=&-Dc\,\theta_c^1-c\,\theta_c^2+b\,\theta_c^4,\nonumber\\
\widetilde{\omega}_{[23]}&=&\theta_c^4,
\end{eqnarray}
with the Weyl form,
\begin{equation}
A=\left (Dc+a_r-2cb\right
)\theta_c^1-2c\,\theta_c^2+b\,\theta_c^4.
\end{equation}
The complete set of Torsion and Curvature forms can be written as
\begin{eqnarray}
T^1&=&0,\nonumber\\
T^2&=&0,\nonumber\\
T^3&=&I_1\theta^1\wedge\theta^4,\nonumber\\
\widetilde{\Omega}_{23}&=&I_2\,\theta_c^2\wedge\theta_c^1+I_3\,\theta_c^3\wedge\theta_c^1,\nonumber\\
\widetilde{\Omega}_{13}&=&I_4 \,\theta^1\wedge\theta^2+I_5  \,\theta^1\wedge\theta^3+I_6 \, \theta^1\wedge\theta^4-I_3  \,\theta^2\wedge\theta^3,\label{Cartan}\\
\widetilde{\Omega}_{12}&=&I_7 \, \theta^1\wedge\theta^2+I_8 \,\theta^1\wedge\theta^3+I_9 \,\theta^1\wedge\theta^4+I_{10} \,\theta^2\wedge\theta^3+I_{11} \,\theta^2\wedge\theta^4,\nonumber\\
\widetilde{\Omega}_{22}&=&\frac{I_8+I_4}{2}
\,\theta^1\wedge\theta^2+2(I_5-I_{10})
\,\theta^1\wedge\theta^3+I_{11} \,\theta^1\wedge\theta^4-2I_3
\,\theta^2\wedge\theta^3.\nonumber
\end{eqnarray}
 This connection was first introduced by Cartan in (\cite{Cartan}), and was called normal metric connection.
 In that work, the steps followed to construct the connection were not clear a priori, only a posteriori one could see
 its intrinsic meaning. Now, with the help of the Cartan equivalence method and its associated invariants (see eq. \ref{Nurowski}), it is
 clear that the normal metric connection is in one to one correspondence with 3rd order ODEs that are equivalent
 under point transformations.

In particular, the vanishing of the invariant $I_1$, known as
Wunschmann invariant~\cite{W}, yields a special class of ODEs that
are related to Conformal Gravity, as is shown in the Null surface
formulation of General
Relativity(\cite{KN,Forni,FKN,FNN,FCN3,SCT}) or to Einstein Weyl
spaces in 3D~\cite{Tod}, as we briefly summarize in the next
section.

\subsection{NSF and Einstein-Weyl spaces in 3D}

Now, we briefly review some special cases from these geometries
associated to third order ODEs.\\
Let us calculate the Lie derivative of $h_c$ in the direction of
$e_s$. If we have a space with a metric connection, it is easily
shown that we can rewrite the Lie derivative as:
\begin{equation}
\pounds_{e_s}h_c=-2A_4h_c+2\eta_{k(i}T^k_{j)4}\theta^i\otimes\theta^j.
\end{equation}
Then, in the case of a normal metric connection presented above,
we have
\begin{equation}
\pounds_{e_s}h_c=-2b\,h_c+I_1\theta^1_c\otimes\theta^1_c,
\end{equation}
and if we restrict the class of ODEs to those that satisfy
\begin{equation}
I_1=F_u+2ab+\frac{da}{ds}=0,
\end{equation} we have a solution space where all lorentzian-metrics $g=(\zeta^{-1})^*h$ in the 1-parameter family
are equivalents to each other, i.e. we can choose the conformal
factor $\Omega=(\zeta^{-1})^*a_3$ as
\begin{equation}
D\Omega=-[(\zeta^{-1})^*b]\,\Omega,
\end{equation}
and then we have that $\widetilde{h}=\Omega^2h$ satisfies
\begin{equation}
\pounds_{e_s}\widetilde{h}=0.
\end{equation}
In this case $G=CO(2,1)$. The differential condition $I_1=0$
provides the kinematics for the Null Surface Formulation of Weyl
spaces. This condition is known as ``metricity condition'' and
$I_1$ is known as Wunschmann invariant~\cite{W}. Solutions
$F(u,w,r,s)$ to the metricity condition allow us to construct a
class of diffeomorphic conformal lorentzian metrics, and solutions
$u=Z(x^a,s)$ to $u'''=F(u,w,r,s)$, have the property that their
level surfaces $Z(x^a,s)=const$, are null surfaces of these
conformal metrics. In particular, we can select a Levi-Civita
connection if we require that $F$ be such that we can find a
function $f$ such that $A_i=grad\,f$,
$i=1,2,3$.\\
Finally, it was shown by Tod~\cite{Tod} following Cartan that if
we require the null surfaces to be totally geodesic then we have
another extra condition (in addition to $I_1=0$) on the ODE such
that from any solution to these conditions we can automatically
construct all Einstein-Weyl spaces. This new condition follows
from
\begin{equation}
e_s\; \lrcorner \;dA=0,
\end{equation} and the condition
reads
\begin{equation}
J(F)=2\frac{d^2\,c}{ds^2}+\frac{d}{ds}\,(b_w)-b_u=0.
\end{equation}
Particular solutions to these ODEs that yield Einstein-Weyl spaces can be found in~\cite{Tod}.\\
In particular, Tod showed that if $F_{rrr}=0$ then $dA=0$, and
from this, we have all conformal Einstein spaces. It is important
to remark that had we used contact transformations instead of
point transformations, then we should have obtained a natural
Cartan Normal conformal connection(\cite{FKNN}). The same result
is obtained in the case of contact transformations if we start
with a pair of second order PDE's, and then we get all conformal
Lorentzian-Weyl geometries in four dimensions~\cite{GKNP}.

\section{Pairs of Partial Differential Equations}
In this section we study the geometry associated with the
following pair of differential equations

\begin{equation}
\begin{split}
Z_{ss} &= S(Z, Z_{s}, Z_{s^{*}}, Z_{ss^{*}}, s, s^{*}), \\
Z_{s^{*}s^{*}} &= S^{*}(Z, Z_{s}, Z_{s^{*}}, Z_{ss^{*}}, s,
s^{*}), \label{diffeqs}
\end{split}
\end{equation}
where $s$ is a complex variable and $S(Z, Z_{s}, Z_{s^{*}},
Z_{ss^{*}}, s, s^{*}) $ satisfies the integrability condition
\begin{equation}
D^{2} S^{*} = D^{*2} S,
\end{equation}
and the weak inequality
\begin{equation}
1-S_{R}S^{*}_{R} > 0.
\end{equation}
The symbols $D, D^*$ in the above expressions denote total
derivatives in the $s$ and $s^{*}$ directions respectively and
their action on an arbitrary function $H=H(Z, W, W^{*}, R, s,
s^{*})$ , is defined as
\begin{align}
\frac{dH}{ds} & \equiv D H \equiv H_{s} + W H_{Z} + S H_{W} + R
H_{W^{*}} + T H_{R}, \label{defD} \\
\frac{dH}{ds^{*}} & \equiv D^{*} H \equiv H_{s^{*}} + W^{*} H_{Z}
+ R H_{W} + S^{*} H_{W^{*}} + T^{*} H_{R}, \label{defDstar}
\end{align}
where
\[
T = D^{*} S, \qquad T^{*} = D S^{*}.
\]
>From the weak  inequality and the Frobenius theorem one can show that the solutions $Z = Z(x^{a},s,s^{*})$ of
(\ref{diffeqs}) depend on four parameters, namely $x^{a}$. Hence the solution space (the space of constants of
integration) is a four-dimensional space.

We now consider the problem of equivalent second order PDEs under
the group of point transformations.

Let ${\bf x}=(Z, Z_{s}, Z_{s^{*}}, Z_{ss^{*}}, s, s^{*}) \equiv
(Z, W, W^{*}, R, s, s^{*})$. As we did in the preceding section,
we identify the spaces $(x^a)\Leftrightarrow (Z,W,W^*,R)$ for any
values of $(s,s^*)$ and we treat this relationship as a coordinate
transformation between the two sets. Their exterior derivatives
\begin{equation}
\begin{split}
dZ &= Z_{a}dx^{a} + Wds + W^{*}ds^{*}, \\
dW &= W_{a}dx^{a} + Sds + Rds^{*}, \\
dW^{*} &= W^{*}_{a}dx^{a} + Rds + S^{*}ds^{*}, \\
dR &= R_{a}dx^{a} + Tds + T^{*}ds^{*},
\end{split}
\end{equation}
can be rewritten as the Pfaffian forms of six  one-forms
\begin{equation}
\begin{split}
\omega^{1} & = dZ - W ds - W^{*} ds^{*},  \\
\omega^{2} & = dW - S ds - R ds^{*},  \\
\omega^{3} & = dW^{*} - R ds - S^{*} ds^{*},  \\
\omega^{4} & = dR - T ds - T^{*} ds^{*}, \\
\omega^5&= ds,\\
\omega^6&=ds^*.
\end{split}
\end{equation}
The vanishing of the four $\omega^{i}, i=1-4$ is equivalent to the
PDEs of Eqs. (\ref{diffeqs}). The point transformation $\bar{{\bf
x}}=\phi({\bf x})$ gives ${\bf \theta}= g \; \omega$, where

$$
 g=\left (\begin{array}{cccccc}
a_1&0&0&0&0&0\\
a_2&a_3&a_4&0&0&0\\
a_2^*&a_4^*&a_3^*&0&0&0\\
a_5&a_6&a_6^*&a_{7}&0&0\\
a_8&0&0&0&a_9&a_{10}\\
a_8^*&0&0&0&a_{10}^*&a_9^*
\end{array}
\right ),
$$
with $a_1, a_5$ and $a_7$ real functions. Differentiating ${\bf
\theta}$, we obtain
\begin{eqnarray}
d{\bf \theta}&=& d  g\;\wedge{\bf \omega}+ g\;d{\bf \omega}\\
&=&d  g\; g^{-1}\wedge  g\;{\bf \omega}+ g\;d{\bf \omega}\\
&=&\Pi\wedge{\bf \theta}+T_{ij} \theta^i\wedge\theta^j,
\end{eqnarray}
where $T_{ij} \theta^i\wedge\theta^j= g\;d{\bf \omega}$,
$$
\Pi=d g\; g^{-1}=\left (\begin{array}{cccccc}
\pi^1&0&0&0&0&0\\
\pi^2&\pi^3&\pi^4&0&0&0\\
\pi^{*2}&\pi^{*4}&\pi^{*3}&0&0&0\\
\pi^5&\pi^6&\pi^{*6}&\pi^{7}&0&0\\
\pi^8&0&0&0&\pi^9&\pi^{10}\\
\pi^{*8}&0&0&0&\pi^{*10}&\pi^{*9}
\end{array}
\right ).
$$

After absorbing some torsion components we obtain

\begin{eqnarray}
d\theta^1&=&\pi^1\wedge\theta^1+I_1\;\theta^3\wedge\theta^6-I_2\;\theta^3\wedge\theta^5+I^*_1\;\theta^2\wedge\theta^5-I^*_2\;\theta^2\wedge\theta^6,\nonumber\\
d\theta^2&=&\pi^2\wedge\theta^1+\pi^3\wedge\theta^2+\pi^4\wedge\theta^3
+I_3\theta^4\wedge\theta^6-I_4\;\theta^4\wedge\theta^5,\nonumber\\
d\theta^3&=&\pi^{*2}\wedge\theta^1+\pi^{*3}\wedge\theta^3+\pi^4\wedge\theta^2
+I^*_3\;\theta^4\wedge\theta^5-I^*_4\;\theta^4\wedge\theta^6,\nonumber\\
d\theta^4&=&\pi^7\wedge\theta^4+\pi^5\wedge\theta^1+\pi^6\wedge\theta^2+\pi^{*6}\wedge\theta^3,\label{theta1_par}\\
d\theta^5&=&\pi^{10}\wedge\theta^6+\pi^{9}\wedge\theta^5+\pi^8\wedge\theta^1,\nonumber\\
d\theta^6&=&\pi^{*10}\wedge\theta^5+\pi^{*9}\wedge\theta^6+\pi^{*8}\wedge\theta^1.\nonumber
\end{eqnarray}
with $$ \displaystyle{\begin{array}{ll}
\displaystyle{I_1=\frac{a_1 (a_4a_{10}+a_3 a_9)}{(a^*_9 a_9-a_{10}
a^*_{10})(a^*_4a_4-a^*_3 a_3)}},
&\displaystyle{I_2=\frac{a_1 (a_4a^*_{9}+a_3 a^*_{10})}{(a^*_9 a_9-a_{10} a^*_{10})(a^*_4a_4-a^*_3 a_3)}},\\\\
\displaystyle{I_3=\frac{a_4a_{10}-a_3 a_9+a_3 a_{10} S_R-a_4 a_{9}
S^*_R}{a_7(a^*_9 a_9-a_{10} a^*_{10})}},
&\displaystyle{I_4=\frac{a_4a^*_{9}-a_3 a^*_{10}+a_3 a^*_{9}
S_R-a_4 a^*_{10} S^*_R}{a_7(a^*_9 a_9-a_{10} a^*_{10})}}.
\end{array}}$$

Normalizing the invariants $I_1=I_3=0$ and $I_2=I_4=1$, the matrix
$ g$ becomes

$$
 g=\left (\begin{array}{cccccc}
a_1&0&0&0&0&0\\
a_2&a_3&a_3 b&0&0&0\\
a_2^*&a_3^* b^*&a_3^*&0&0&0\\
a_5&a_6&a_6^*&\frac{a^*_3 a_3}{\alpha^2\, a_1}&0&0\\
a_8&0&0&0&\frac{-b a_1}{a^*_3 (1-b b^*)}&\frac{a_1}{a^*_3 (1-b b^*)}\\
a_8^*&0&0&0&\frac{a_1}{a_3 (1-b b^*)}&\frac{-b^* a_1}{a_3 (1-b
b^*)}
\end{array}
\right ),
$$
where $\displaystyle{b=\frac{\sqrt{1-S_R S^*_R}-1}{S^*_R}}$  and $\displaystyle{\alpha^2=\frac{1+b b^*}{(1-b
b^*)^2}}$.\\

Thus, $\pi^4$, $\pi^7$, $\pi^9$ and $\pi^{10}$ are functionals of $a_1$, $a_2$, $a_3$, and $S$.

After absorbing some torsion coefficients, the second loop gives
the following structure equations

\begin{eqnarray}
d\theta^1&=&\pi^1\wedge\theta^1-\theta^3\wedge\theta^5-\theta^2\wedge\theta^6,\nonumber\\
d\theta^2&=&\pi^2\wedge\theta^1+\pi^3\wedge\theta^2-\theta^4\wedge\theta^5+I_5\,\theta^4\wedge\theta^3
+I_6\,\theta^3\wedge\theta^5+I_7\;\theta^3\wedge\theta^6,\nonumber\\
d\theta^3&=&\pi^{*2}\wedge\theta^1+\pi^{*3}\wedge\theta^3-\theta^4\wedge\theta^6+I^*_5\theta^4\wedge\theta^2
+I^*_6\;\theta^2\wedge\theta^6+I^*_7\;\theta^2\wedge\theta^5,\nonumber\\
d\theta^4&=&-\pi^1\wedge\theta^4+\pi^3\wedge\theta^4+\pi^{*3}\wedge\theta^4+\pi^5\wedge\theta^1+\pi^6\wedge\theta^2+\pi^{*6}\wedge\theta^3\nonumber\\&&+2I^*_6
\,\theta^5\wedge\theta^4+2I_6 \,\theta^6\wedge\theta^4,\nonumber\\
d\theta^5&=&\pi^{1}\wedge\theta^5-\pi^{*3}\wedge\theta^5+\pi^8\wedge\theta^1+I_5\,\theta^4\wedge\theta^6
+I_8\,\theta^2\wedge\theta^5+I^*_8\,\theta^2\wedge\theta^6\nonumber\\&&+I_9\,\theta^3\wedge\theta^6,\nonumber\\
d\theta^6&=&\pi^{1}\wedge\theta^6-\pi^{3}\wedge\theta^6+\pi^{*8}\wedge\theta^1+I^*_5\,\theta^4\wedge\theta^5+I^*_8\,\theta^3\wedge\theta^6+I_8\,\theta^3\wedge\theta^5\nonumber\\&&+I^*_9\,\theta^2\wedge\theta^5.\label{theta1_par2}
\end{eqnarray}

The invariants are
\begin{eqnarray*}
I_5&=&\frac{a_1 \alpha^2 b_R}{(a^*_3)^2 (1-b b^*)},\\
I_6&=&\frac{1}{1-b b^*}\left(\frac{a^*_6}{a^*_3}-\frac{a_6 b}{a_3^*}-\frac{a_2(1-b b^*)}{a_1}-\frac{a_3 (a^*-ab)}{a_1}\right),\\
I_7&=& -\frac{a_3^2\,\mathcal{W}}{  a_1 a^*_3},\\
I_8&=&\frac{a_2\,\alpha^2 b^*_R }{(a_3)^2\,(1-b b^*)}+\frac{b^*_{W} b-b^*_{W^*}}{a_3\,(1-b b^*)^2}+ \frac{\alpha^2 b^*_R (a^*-ab)}{a_3\,(1-b b^*)^2}-\frac{a_{8}^*}{a_1},\\
I_9&=&\frac{a_2\alpha^2 b_R}{(a^*_3)^2\,(1-b b^*)}+\frac{b_W
b-b_{W^*}}{(1-b b^*)^2}+\frac{a_3\alpha^2 b_R (a^*-ab) }{a_3^*(1-b
b^*)^2},
\end{eqnarray*}
where
\begin{eqnarray*}
a&=&b^* S_{W}+b S^*_{W}+\frac{Db^*+b D^*b^*+(b^*)^2Db+b^*D^* b}{1-b b^*},\\
\mathcal{W}&=&\frac{1}{1-b b^*}\left(Db - b S_W +S_{W^*}+b(D^* b+b S^*_{W^*}-b^2S^*_W)\right).
\end{eqnarray*}

The expression $\mathcal{W}$ is known as the generalized W\"unschmann invariant.

Choosing $I_6=I_8=0$ we obtain
\begin{eqnarray*}
a_6&=&\frac{a a_3 a^*_3+a_3 a^*_2+b^* a^*_3 a_2}{a_1},\\
a_{8}&=&\frac{a_1 a^*_2\,\alpha^2 b_R }{(a^*_3)^2(1-b b^*)} +
\frac{a_1(b_{W^*} b^*-b_{W})}{a^*_3\,(1-b b^*)^2}+
\frac{a_1\,\alpha^2 b_R (a-a^*b^*)}{a^*_3\,(1-b b^*)^2}.
\end{eqnarray*}

Then the structure equations can be written as
\begin{eqnarray}
d\theta^1&=&\pi^1\wedge\theta^1-\theta^3\wedge\theta^5-\theta^2\wedge\theta^6,\nonumber\\
d\theta^2&=&\pi^2\wedge\theta^1+\pi^3\wedge\theta^2-\theta^4\wedge\theta^5+I_5\,\theta^4\wedge\theta^3
+I_7\;\theta^3\wedge\theta^6,\nonumber\\
d\theta^3&=&\pi^{*2}\wedge\theta^1+\pi^{*3}\wedge\theta^3-\theta^4\wedge\theta^6+I^*_5\theta^4\wedge\theta^2
+I^*_7\;\theta^2\wedge\theta^5,\label{theta3_par}\\
d\theta^4&=&-\pi^1\wedge\theta^4+\pi^3\wedge\theta^4+\pi^{*3}\wedge\theta^4+\pi^5\wedge\theta^1+\pi^6\wedge\theta^2+\pi^{*6}\wedge\theta^3,\nonumber\\
d\theta^5&=&\pi^{1}\wedge\theta^5-\pi^{*3}\wedge\theta^5+\pi^8\wedge\theta^1+I_5\,\theta^4\wedge\theta^6
+I_9\,\theta^3\wedge\theta^6,\nonumber\\
d\theta^6&=&\pi^{1}\wedge\theta^6-\pi^{3}\wedge\theta^6+\pi^{*8}\wedge\theta^1+I^*_5\,\theta^4\wedge\theta^5+I^*_9\,\theta^2\wedge\theta^5.\nonumber
\end{eqnarray}
Replacing $a_6$ and $a_8$ in the matrix $ g$ fixes $\pi^6 $ and
$\pi^8$.

Computing again  the  third loop and after the absorption of the
torsion gives the following structure equations for $d\theta^i$,
$i=1...4$ (the remaining exterior derivatives have been omitted
since they are not used to obtain the null tetrad and torsion
tensor),

\begin{eqnarray}
d\theta^1&=&\pi^1\wedge\theta^1-\theta^3\wedge\theta^5-\theta^2\wedge\theta^6,\nonumber\\
d\theta^2&=&\pi^2\wedge\theta^1+\pi^3\wedge\theta^2-\theta^4\wedge\theta^5+I_5\,\theta^4\wedge\theta^3
+I_7\;\theta^3\wedge\theta^6,\nonumber\\
d\theta^3&=&\pi^{*2}\wedge\theta^1+\pi^{*3}\wedge\theta^3-\theta^4\wedge\theta^6+I^*_5\theta^4\wedge\theta^2
+I^*_7\;\theta^2\wedge\theta^5,\label{theta3final}\\
d\theta^4&=&-\pi^1\wedge\theta^4+\pi^2\wedge\theta^2+\pi^{*2}\wedge\theta^3+\pi^3\wedge\theta^4+\pi^{*3}\wedge\theta^4+\pi^5\wedge\theta^1\nonumber\\
&&+I_{10} \,\theta^2\wedge\theta^6+I^*_{10}
\,\theta^3\wedge\theta^5 +I_{11} \,\theta^3\wedge\theta^6+I^*_{11}
\,\theta^2\wedge\theta^5,\nonumber
\end{eqnarray}
where
\begin{eqnarray}
I_{10}&=&2\left(\frac{a_5}{a_1}-\frac{a_2a^*_2}{a^2_1}-\frac{a_3a^*_3}{\alpha^2a_1^2}\,c\right)+i\,\text{Im}[I_{10}],\label{a5}\\
I_{11}&=& \frac{a^*_3 a_3^2}{a_1^2}\frac{(1-b b^*)}{(1+b
b^*)}(b^2\Delta^*-\Delta)-2\mathcal{W}\frac{a_3^2}{a_1^2}\left( a_2^*-\frac{a^*_3 b^* (a^*-ab)}{1-(b
b^*)^2}\right), \label{funW}
\end{eqnarray}
with c a real function
\begin{eqnarray}
c&=&-\frac{Da+D^{*}a^{*}+T_{W}+T_{W^{*}}^{*}}{4}-\frac{
aa^{*}(1+6bb^{*}+b^{2}b^{*2})}{2(1+bb^{*})^{2}}  \label{eqc} \nonumber\\
&&+\frac{(1+bb^{*})(bS_{Z}^{*}+b^{*}S_{Z})}{2(1-bb^{*})^{2}}+\frac{
a(2ab-b^{*}S_{W^{*}})+a^{*}(2a^{*}b^{*}-bS_{W}^{*})}{2(1+bb^{*})}\nonumber\\
&&+\alpha^4\left[2b^*\left(a-a^*b^*\right)\mathcal{W}+2b\left(a^*-ab\right)\mathcal{W}^*\right],
\end{eqnarray}
and $\text{Im}[I_{10}]$, the imaginary part of $I_{10}$,
\begin{eqnarray}
i\,\text{Im}[I_{10}]&=&\frac{a_3a^*_3\alpha^2}{a^2_1}\left [2b^*\left(a-a^*b^*\right)\mathcal{W}-
2b\left(a^*-ab\right)\mathcal{W}^*\label{ima5}\right .\\
&&+\left
.\left(1-bb^*\right)^2(\Gamma-\Gamma^*)-2\left(1-(bb^*)^2\right)\left(b^*\Delta-b\Delta^*\right)\right
],
\end{eqnarray}
and where $\Delta$ and $\Gamma$ are explicit function of $W$,
$W^*$ and its derivatives, namely

\begin{eqnarray*}
\Delta&=&-\alpha^2 b \mathcal{W} \mathcal{W}^* +2(1+bb^{*})^{-1}\left \{\rho (1-bb^{*})+\nu
[a^{*}b^{*}+a(1-bb^{*}-b^{2}b^{*2})]\right . \\
&&+ \left .b^{2}\rho ^{*}(1-bb^{*})+b^{2}\nu
^{*}[ab+a^{*}(1-bb^{*}-b^{2}b^{*2})]\right\}\\
\Gamma&=&-b^{*}\left [4\rho +2\nu (2abb^{*}+a^{*}b^{*}+3a)\right
],
\end{eqnarray*}
with
\begin{eqnarray*}
\rho &= &\frac{\alpha^2}{2(1-bb^{*})}\left \{b^{*}D\mathcal{W}+D^{*}\mathcal{W}
+\mathcal{W}(b^{*2}S_{W^{*}}-2b^{*}S_{W}-2bS_{W}^{*}+S_{W^{*}}^{*})\right\}, \\
\nu &=&\frac{\mathcal{W}}{1-(bb^{*})^{2}}.
\end{eqnarray*}

>From (\ref{a5}) and taking into account that $a_5$ must be a real parameter, we can solve for $a_5$ demanding
the vanishing of the real part of $I_{10}$, i.e., by doing $\text{Re}\left[I_{10}\right]=0$. We find that
\begin{equation}
a_5=\frac{1}{a_1}\left(a_2a^*_2+\frac{a_3a^*_3}{\alpha^2}\,c\right).
\end{equation}\\
\textbf{Remark 4} \textit{Note that with this choice  of  $a_5$,
we have that  $I_{10}$ and $I_{11}$ depend only
on $\mathcal{W}$, $\mathcal{W}^*$ and its derivatives.}\\\\
\textbf{Remark 5} \textit{After the third loop we have fixed some
parameters and others remain free. The free parameters are
$a_1,a_2,a^*_2,a_3,a^*_3$, and the fixed ones are:}
\begin{eqnarray}
a_4&=&ba_3,\nonumber\\
a_5&=&\frac{1}{a_1}\left(a_2a^*_2+\frac{a_3a^*_3}{\alpha^2}\,c\right),\nonumber\\
a_6&=&\frac{a a_3 a^*_3+\alpha^2\left(a_3 a_2+b^* a^*_3
a_2\right)}{\alpha^2a_1},\nonumber\\
a_7&=&\frac{a_3a^*_3}{\alpha^2a_1},\nonumber\\
a_{8}&=&\frac{a_1 a^*_2\,\alpha^2 b_R }{(a^*_3)^2(1-b b^*)} +
\frac{a_1(b_{W^*} b^*-b_{W})}{a^*_3\,(1-b b^*)^2}+
\frac{a_1\,\alpha^2 b_R (a-a^*b^*)}{a^*_3\,(1-b b^*)^2}\nonumber\\
a_9&=&\frac{-b a_1}{a^*_3 (1-b b^*)},\nonumber\\
a_{10}&=&\frac{a_1}{a^*_3 (1-b b^*)}.
\end{eqnarray}

We could continue applying Cartan's equivalence method, prolonging
and  studying the different branches, but for
our purposes it will not be needed.\\\\
\textbf{Remark 6} \textit{It is important to remark that with Cartan's method we have showed that $\mathcal{W}$
is a relative invariant under point transformations (a subset of the contact transformations). It also follows
from our construction that $\mathcal{W}$ is a relative invariant under contact transformation since the proof is
identical up
to the loop presented in this work. (For a different proof see \cite{FKN}).}\\

Note that $a_1$ is a real parameter, whereas $a_2$ and $a_3$ are complex. Their geometrical meaning can be
emphasized if they are rewritten as
$$
a_1=\mu,\; \; a_2=\Omega\,\gamma \,e^{i\psi},\; \; a_3=\Omega\,\alpha\, e^{i\psi}.
$$
with real parameters $\mu$, $\Omega$ and $\psi$ and a complex
$\gamma$.

\section{Null coframes} Following the
last section, we write
$$
a_1=\mu,\; \; a_2=\Omega\,\gamma \,e^{i\psi},\; \; a_3=\Omega\,\alpha\, e^{i\psi},\; \; a_4=\Omega\,\alpha\,
e^{i\psi}b,$$
$$
a_5=\frac{\Omega^2}{\mu}\left(\gamma\gamma^*+c\right),\; \; a_6=\frac{\Omega^2}{\mu}\left(\alpha\gamma+\alpha
b^*\gamma+a\right),\; \; a_7=\frac{\Omega^2}{\mu}.
$$
Defining
\begin{eqnarray}
\theta_c^{1}\, &=& \omega ^{1}, \\
\theta_c^{2} &=&\alpha (\omega ^{2}+b\omega ^{3}),\\
\theta_c^{3} &=&\alpha (\omega ^{3}+ b^{*}\omega ^{2}),
\nonumber \\
\theta_c^{4}\, &=&\omega ^{4}+a\omega ^{2}+a^{*}\omega ^{3}+
c\,\omega ^{1}.
\end{eqnarray}
we can write the forms $\theta^i$ in a similar manner as in  3D,
\begin{eqnarray}
\theta^{1} &=&\mu\theta_c^{1}, \\
\theta^{2} &=&\Omega e^{i\psi}\left(\theta_c^{2}+\gamma \theta_c^{1}\right), \\
\theta^{3} &=&\Omega e^{-i\psi}\left(\theta_c^{3}+\gamma
^{*}\theta_c^{1}\right), \\
\theta^{4}&=&\frac{\Omega^2}{\mu}\left(\theta_c^{4}+\gamma\theta_c^{3}+\gamma
^{*}\theta_c^{2}+\gamma \gamma ^{*}\theta_c^{1}\right).
\end{eqnarray}
We define the quadratic  form over $J^2(\mathbb{R}^2,\mathbb{R})$,
\begin{equation}
h(\textbf{x})=2\theta^{(1}\otimes\theta^{4)}-2\theta^{(2}\otimes\theta^{3)}=\eta_{ij}\theta^i\otimes\theta^j,
\end{equation}
with
\begin{equation}
\eta_{ij}=\left(%
\begin{array}{cccc}
  0& 1 & 0 & 0 \\
  -1 & 0 & 0 & 0 \\
  0& 0 & 0 & 1 \\
  0& 0 & 1 & 0 \\
\end{array}%
\right).
\end{equation}
This quadratic form induces a two-parameter family of lorentzian
conformal metrics in the solution space M. The
forms $\theta^i$ yield a null tetrad, and the parameters $\Omega,\mu,\psi$ and $\gamma$ acquire geometrical meaning:\\\\
$\bullet$\; $\mu$  plays the role of a  boost in the direction of the null vector $e_1$ dual to
$\theta^1$.\\\\
$\bullet$\; $\gamma$ and $\gamma^*$ parametrize a null rotation
around $e_1$.\\\\
$\bullet$\; $e^{i\psi}$  is a  space-like rotation around a fixed axis
in the celestial sphere.\\\\
$\bullet$\; $\Omega$ is a conformal factor applied to the tetrad.\\\\
\textbf{Remark 7} \textit{Using the definition of the quadratic
form together with the exterior derivatives of the $\theta^i$ in
eq. (125) it is straightforward to show that
\begin{equation}
\pounds_{e_s}h\propto h+F[\mathcal{W},\mathcal{W}^*],\;\;\; \pounds_{e_s^*}h\propto
h+F^*[\mathcal{W},\mathcal{W}^*],
\end{equation}
where the tensor F is a functional of $\mathcal{W}$ and its derivatives that vanishes when $W=0$. Then if
$\mathcal{W}=\mathcal{W}^*=0$, $G$ becomes the conformal group $CO(3,1)$, with $s$ and $s^*$ space-like rotation
parameters. All the metrics in the
family are conformal to each other.}\\

In an equivalent way to the three dimensional problem we can introduce a connection and associated torsion that
satisfy requirements I, II and III. It is straightforward to show that the Torsion has the form
\begin{eqnarray}
T^1&=&0,\nonumber\\
T^2&=&I_7\;\theta^3\wedge\theta^6,\nonumber\\
T^3&=&I^*_7\;\theta^2\wedge\theta^5,\nonumber\\
T^4&=&I_{10} \,\theta^2\wedge\theta^6+I^*_{10}
\,\theta^3\wedge\theta^5 +I_{11} \,\theta^3\wedge\theta^6+I^*_{11}
\,\theta^2\wedge\theta^5.\label{torsion}
\end{eqnarray}\\
\textbf{Remark 8} \textit{It follows from the above equations that the vanishing of $\mathcal{W}$, the
generalized W\"unschmann invariant, gives a torsion free connection. Moreover, using remark 7 we recover
NSF.}\\

Note that, as in the 3-dim case, the skew part of the connection is completely determined from (\ref{torsion}).
We could, in principle, also determine the Weyl part of the connection following a similar procedure as in the
3-dim case. Associated with the five undetermined $a_i$ we have the corresponding $\pi_i$. The seven two-forms
$(d\pi_i, d\theta_5, d\theta_6)$ can be used to construct the seven curvature components $(\Omega_{[ij]}, dA)$.
In this case, the algebraic restriction on the curvature components arise from the invariants contained in
$(d\theta_5, d\theta_6)$ in an equivalent way to eq.(\ref{curvatura}). Assuming this program can be completed we
would then have the result that a normal metric connection in 4-dim is in one to one correspondence with pairs
of PDEs of the form (\ref{diffeqs}) that are equivalent under point transformations.

Note also that if we consider the problem of contact transformations we should end up with a higher dimensional
group that includes the translations. In this last case the spatial part of the Weyl form will not be determined
by our construction.

\section{Conclusions}
In this work we have used the powerful method developed by Cartan to study the equivalence problem  of third
order ODEs and a pair of second order of PDEs under point transformations. We obtained from first principles the
null tetrads, the conformal metric and the so called normal metric connection associated with these equations.
These results complement and give a more clear understanding about the origin of the conformal geometry
underlying these differential equations.

In particular, it follows from this construction that every conformal geometry of a space-time in three or four
dimensions is contained in a subclass of differential equations defined by the vanishing of a relative invariant
know as the W\"unschmann invariant. In this particular case the tetrads associated to the two PDEs are related
by a conformal transformation and thus, the conformal structure is unique.

It is interesting and surprising at the same time to see that general relativity, or more precisely conformal
gravity, is  contained in a special subclass of differential equations, i.e., those with a vanishing
W\"unschmann invariant. Note that from this point of view the space-time emerges as the solution space of the
differential equations. Clearly this is a non standard prescription of general relativity.

One might ask what is the physical meaning of these starting differential equations on a fiducial space. At
least in NSF we know the answer. The intersection of a future light cone from a point of an asymptotically flat
space-time with the null boundary is called a light cone cut of null infinity. It can be shown that this cut
satisfies the differential equations presented in this work and the points of the fiducial space are the points
of the null boundary \cite{KN}. Using these results we can easily show that solutions of the light cone cut
equation defined up to point or contact transformations on the null boundary yield a unique conformal structure
on the space-time.

 Finally, it would be very interesting to investigate all branches of the equivalence problem associated to these
 equations. In this way we could find all symmetries of these ODEs and PDEs, and we have at our disposal a powerful
 technique to generate new solutions to these equations from given known solutions.\\

\section{Acknowledgments}
E.G. M.I. and C.N.K. would like to thank CONICET for support.

\end{document}